\documentclass{article}

\usepackage[english]{babel}

\usepackage[a4paper,top=2cm,bottom=2cm,left=3cm,right=3cm,marginparwidth=1.75cm]{geometry}

\usepackage{amssymb}
\usepackage{siunitx}
\PassOptionsToPackage{hyphens}{url}\usepackage{hyperref}
\usepackage{cleveref}
\usepackage[utf8]{inputenc}
\usepackage[right]{lineno}
\usepackage{csquotes}
\usepackage{booktabs}
\usepackage{longtable}
\usepackage{adjustbox}
\usepackage{array}
\usepackage{url}
\usepackage{titlesec}
\usepackage{authblk}
\usepackage{xcolor} % Load the xcolor package for color options
\usepackage{comment}

\newcommand{\sqrtsnn}{\sqrt{s_{_\text{NN}}}}

\titleformat{\subsection}
  {\mdseries\itshape\large} % Medium series, italic shape, and large font size
  {\thesubsection}{1em}{} % Numbering, spacing, and the section title itself

\usepackage[english]{babel}
\usepackage[style=authoryear,backend=biber,natbib=true,maxcitenames=2,uniquelist=false]{biblatex}
\DeclareNameAlias{sortname}{family-given}
\DeclareNameAlias{default}{family-given}

\renewbibmacro{in:}{}
\DeclareFieldFormat[article]{title}{\mkbibquote{#1}\addcomma}
\DeclareFieldFormat[book]{title}{\mkbibemph{#1}\addcomma}
\DeclareFieldFormat[bookinbook]{title}{\mkbibemph{#1}\addcomma}
\DeclareFieldFormat[inbook]{title}{\mkbibquote{#1}\addcomma}
\DeclareFieldFormat[incollection]{title}{\mkbibquote{#1}\addcomma}
\DeclareFieldFormat[inproceedings]{title}{\mkbibquote{#1}\addcomma}
\DeclareFieldFormat[manual]{title}{\mkbibemph{#1}\addcomma}
\DeclareFieldFormat[misc]{title}{\mkbibemph{#1}\addcomma}
\DeclareFieldFormat[thesis]{title}{\mkbibemph{#1}\addcomma}
\DeclareFieldFormat[unpublished]{title}{\mkbibquote{#1}\addcomma}
\DeclareFieldFormat[patent]{title}{\mkbibemph{#1}\addcomma}
\DeclareFieldFormat[report]{title}{\mkbibemph{#1}\addcomma}
\DeclareFieldFormat[online]{title}{\mkbibquote{#1}\addcomma}
\DeclareFieldFormat[software]{title}{\mkbibemph{#1}\addcomma}
\DeclareFieldFormat[booklet]{title}{\mkbibemph{#1}\addcomma}
\DeclareFieldFormat[periodical]{title}{\mkbibemph{#1}\addcomma}
\DeclareFieldFormat[standard]{title}{\mkbibemph{#1}\addcomma}

\DeclareFieldFormat[article]{journaltitle}{\iffieldundef{shortjournal}{\mkbibemph{#1}\addcomma}{\mkbibemph{\printfield{shortjournal}}\addcomma}}
\DeclareFieldFormat{volume}{\bibstring{volume}~#1}
\DeclareFieldFormat{number}{\bibstring{number}~#1}

\DefineBibliographyStrings{english}{
  volume = {Vol.},
  number = {No.}
}

\renewbibmacro*{volume+number+eid}{%
  \printfield{volume}%
  \setunit*{\addspace}%
  \printfield{number}%
  \setunit{\addcomma\space}%
  \printfield{eid}}

\renewbibmacro*{journal+issuetitle}{%
  \usebibmacro{journal}%
  \setunit*{\addcomma\space}%
  \usebibmacro{volume+number+eid}%
  \setunit{\addcomma\space}%
  \usebibmacro{issue+date}}

\renewbibmacro*{publisher+location+date}{%
  \printlist{publisher}%
  \iflistundef{location}
    {\setunit*{\addcomma\space}}
    {\setunit*{\addcolon\space}}%
  \printlist{location}%
  \setunit*{\addcomma\space}%
  \usebibmacro{date}}

\DeclareCiteCommand{\cite}[\mkbibparens]
  {\usebibmacro{prenote}}
  {\usebibmacro{citeindex}%
   \usebibmacro{cite}}
  {\multicitedelim}
  {\usebibmacro{postnote}}

\renewbibmacro*{cite:labelyear+extrayear}{%
  \iffieldundef{labelyear}
    {}
    {\printtext[bibhyperref]{%
       \printfield{labelyear}%
       \printfield{extrayear}}}}

\renewbibmacro*{cite:labeldate+extradate}{%
  \iffieldundef{labelyear}
    {}
    {\printtext[bibhyperref]{%
       \printfield{labelyear}%
       \printfield{extradate}}}}

\AtEveryBibitem{
  \clearfield{month}
  \clearfield{day}
  \ifentrytype{book}{
    \clearlist{location}
  }{}
}

\DefineBibliographyStrings{english}{
  andothers = {\textit{et al.},}
}

\DeclareFieldFormat[article]{volume}{\bibstring{jourvol}\addnbspace #1}
\DeclareFieldFormat[article]{number}{\bibstring{number}\addnbspace #1}
\DeclareFieldFormat[article]{volume}{Vol. #1}
\DeclareFieldFormat[article]{number}{No. #1}
\DeclareFieldFormat{url}{\bibstring{available at}\addcolon\space\url{#1}}
\DeclareFieldFormat{urldate}{\mkbibparens{accessed \addspace#1}}

\DeclareFieldFormat{urldate}{%
  \mkbibparens{accessed\space%
    \thefield{urlday}\addspace%
    \mkbibmonth{\thefield{urlmonth}}\addspace%
    \thefield{urlyear}}}

\crefformat{figure}{#2Figure~#1#3}
\Crefformat{figure}{#2Figure~#1#3}
\crefformat{table}{#2Table~#1#3}
\Crefformat{table}{#2Table~#1#3}
\crefformat{section}{#2Section~#1#3}
\Crefformat{section}{#2Section~#1#3}

\author[2]{Carolina~ARATA}
\author[1]{Fran\c{c}ois~ARLEO}
\author[2]{Benjamin~AUDURIER}
\author[2]{Alberto~BALDISSERI}
\author[6]{Nicole~BASTID}
\author[1]{Guillaume~BATIGNE}
\author[3]{Iouri~BELIKOV}
\author[1]{Marcus~BLUHM}
\author[2]{Francesco~BOSSU}
\author[2]{Hervé~BOREL}
\author[2]{Javier~CASTILLO~CASTELLANOS}
\author[1]{Paul~CAUCAL}
\author[8]{Cvetan~CHESHKOV}
\author[4]{Gustavo~CONESA~BALBASTRE}
\author[5]{Zaida~CONESA~DEL~VALLE}
\author[1]{Maurice~COQUET}
\author[2]{Imanol~CORREDOIRA~FERNANDEZ}
\author[6]{Philippe~CROCHET}
\author[5]{Bruno~ESPAGNON}
\author[4]{Julien~FAIVRE}
\author[2]{Andrea~FERRERO}
\author[2]{Audrey~FRANCISCO}
\author[7]{Frédéric~FLEURET}
\author[5]{Chris~FLETT}
\author[4]{Christophe~FURGET}
\author[1]{Marie~GERMAIN}
\author[1]{Pol~Bernard~GOSSIAUX}
\author[4]{Rachid~GUERNANE}
\author[1]{Maxime~GUILBAUD}
\author[1]{Manuel~GUITTIERE}
\author[5]{Cynthia~HADJIDAKIS}
\author[3]{Boris~HIPPOLYTE}
\author[3]{Christian~KUHN}
\author[5]{Jean-Philippe~LANSBERG}
\author[6]{Xavier~LOPEZ}
\author[3]{Antonin~MAIRE}
\author[5]{Dukhishyam~MALLICK}
\author[9]{Cyrille~MARQUET}
\author[1]{Gin\'es~MARTINEZ-GARCIA}
\author[5]{Laure~MASSACRIER}
\author[7]{Kara~MATTIOLI}
\author[7]{\'Emilie~MAURICE}
\author[5]{Carlos~MU\~NOZ~CAMACHO}
\author[1]{Marlene~NAHRGANG}
\author[5]{Maxim~NEFEDOV}
\author[7]{\'Elisabeth~NIEL}
\author[5]{Melih~A.~OZCELIK}
\author[2]{Stefano~PANEBIANCO}
\author[1]{Philippe~PILLOT}
\author[18]{Bernard~PIRE}
\author[6]{Sarah~PORTEBOEUF~HOUSSAIS}
\author[2]{Andry~RAKOTOZAFINDRABE}
\author[8]{Niveditha~RAMASUBRAMANIAN}
\author[5]{Patrick~ROBBE}
\author[5]{Hagop~SAZDJIAN}
\author[3]{Serhiy~SENYUKOV}
\author[5]{Christophe~SUIRE}
\author[8]{Antonio~URAS}
\author[5]{Samuel~WALLON}
\author[2]{Michael~WINN}

\affil[1]{Subatech, IMT Atlantique, Nantes-Universit\'e, CNRS/IN2P3, Nantes, France}
\affil[2]{Universit\'e Paris-Saclay, Centre d’Etudes de Saclay (CEA), IRFU, D\'epartment de Physique Nucl\'eaire (DPhN), Saclay, France}
\affil[3]{IPHC, CNRS/IN2P3, Université de Strasbourg, Strasbourg, France}
\affil[4]{LPSC, CNRS/IN2P3, Université de Grenoble, Grenoble, France}
\affil[5]{Universit\'e Paris-Saclay, CNRS/IN2P3, IJCLab, Orsay, France}
\affil[6]{Université Clermont-Auvergne, CNRS, LPCA, 63000 Clermont-Ferrand, France}
\affil[7]{Laboratoire Leprince-Ringuet, CNRS/IN2P3, \'Ecole polytechnique, Palaiseau, France}
\affil[8]{Université de Lyon, CNRS/IN2P3, Institut de Physique des 2 Infinis de Lyon, Lyon, France}
\affil[9]{CPHT, CNRS, Ecole Polytechnique, Institut Polytechnique de Paris, Palaiseau, France}

\title{Prospective report of the French QCD community to the ESPPU 2025 with respect to the program of the LHC Run 5 and beyond and future colliders at CERN}

\begin{document}
\maketitle

\begin{abstract}
This document summarizes the prospective physics plans of the French QCD and Heavy-Ion community, including the experimental programs at the LHC Run 5 and beyond and future colliders at CERN, within the context of the French contribution to the update of the European Strategy in Particle Physics (ESPPU 2025), as discussed in the \href{https://indico.in2p3.fr/event/33460/}{workshop on European Strategy for Particle Physics Update 2025} organised by the \href{https://gdrqcd.in2p3.fr/}{QCD GdR} in Oct.~2024.
\\
\\
%add 6 keywords
\textbf{Keywords:} QCD, Heavy Ions, Quark Gluon Plasma, LHC, FCC
\end{abstract}
%\linenumbers

\section*{Introduction} %\footnote{The following to text motivating the pursue of the exploration of the hadronic matter phase diagram is taken from the reference \cite{ALICE:2022wwr}}}

The Standard Model (SM) of particle physics describes the fundamental constituents of matter and the laws governing their interactions. It also accounts for how collective phenomena and equilibrium properties of matter arise from elementary interactions. Theory makes quantitative statements about the equation of state of SM matter, about the nature of the electroweak and strong phase transitions, and about fundamental properties such as transport coefficients and relaxation times of many-body systems of elementary particles interacting collectively. In addition, there has been considerable theoretical progress in describing how out-of-equilibrium evolution drives non-Abelian matter towards equilibrium. Collisions of nuclei at ultra-relativistic energies offer a unique possibility for testing some key facets of the rich high-temperature thermodynamics of the SM in laboratory-based experiments. They test the strong interaction sector of the SM at energy densities at which partonic degrees of freedom dominate the equilibration processes. They thus give access to the partonic dynamics that drives fundamental non-Abelian matter towards equilibrium and that determines the properties of the high temperature phase of quantum chromodynamics (QCD), the quark–gluon plasma (QGP). [This paragraph is in good part from~\cite{ALICE:2022wwr}. See also \cite{Busza:2018rrf, Apolinario:2022vzg}.] Improving our understanding of the collective behaviour of non-Abelian quantum field theories (of which the QCD and EW sectors of the SM are paramount examples, although only the former is open to experimental study at colliders) has also obvious implications for our understanding of the space-time evolution and phase transitions of the early universe.

The field of ultra-relativistic nuclear collisions has seen enormous progress since its inception in the mid-eighties, from the first signals of colour deconfinement at the SPS to the evidence, at RHIC, for a strongly-coupled QCD medium that quenches hard partons. Nuclear collisions at the LHC offer an ideal environment for a broad program of characterisation of the properties of this unique state of matter. Besides providing access to the highest-temperature, longest-lived experimentally accessible QCD medium, they also offer an abundant supply of self-calibrating heavy-flavour probes. In addition, the very low net baryon density eases significantly the quantitative connection between experimental measurements and lattice QCD calculations.

\section*{Heavy Ion Run at LHC Run 5 and beyond}
\label{sec:HeavyIons}

Full completion of the LHC heavy ion program will require running heavy ion collisions at the LHC until the end of the LHC operation. LHC experiments will pursue the exploration of the strongly interacting matter phase diagram (QCD matter), studying the properties of the ephemeral droplets of the plasma of quarks and gluons (QGP) that have been produced at the LHC since 2010, with better precision, new observables (like multi-charm hadrons, high precision beauty, P-waves quarkonia, ...) and varying collisions systems. In particular Run 5 will be crucial in understanding the onset of collectivity in small systems~\cite{Grosse-Oetringhaus:2024bwr} and its evolution with the size of the system. Classic QGP physics has established a convincing paradigm describing key features of heavy-ion collisions as flow, energy loss, and quarkonium physics at colliders as due to the formation of a strongly coupled QGP. However, this picture remains to a certain extent qualitative. An understanding of the inner workings of the emergence of QGP, its properties and its transition to hadrons, i.e.\ the microscopic picture behind the initial state, the thermalisation,  deconfinement, hadronisation and chiral restoration will be enabled by  high precision beauty quark hadron measurements, charm-charm correlations, multi-charm baryon,  P-wave quarkonia production, dilepton and photon measurements. Many different small and intermediate systems will have to be studied at the LHC, from high multiplicity pp until most central Pb-Pb, scanning well-selected species that can be accelerated in the LHC with various systems : p-A and AA. If in the past Runs 1 and 2, the LHC demonstrated the feasibility and interest of the community for collisions with O, Ar and Xe ions, other species could be considered, depending on their availability at the LHC and the priority established by the community. Further discussions with LHC machine experts, LHC experiment collaborations and within the international heavy ion community will be needed to define precisely the planning, the collision systems to be studied and the associated integrated luminosity. Moreover, intermediate systems will bring the possibility to improve the signal to background ratio, as compared to central Pb-Pb collisions and, of interest for nuclear physics, to study the nuclear structure of the accelerated nucleus, for spherical nucleus and for nucleus exhibiting exotic nuclear structure, see for a current overview of activities at the light ions workshop at CERN in November 2024~\href{https://indico.cern.ch/event/1436085/}{https://indico.cern.ch/event/1436085/}.

In addition, the Run 5 will allow to complete the fixed target LHC programme, led by the LHCb collaboration \cite{2707819}, to perform a precision heavy-flavour and quarkonium program from pp up to large nuclear systems from midrapidity close to the kinematic edge. Moreover, the comparatively large kinematic coverage with full instrumentation allows for unique measurements also in the soft physics programme and compared to RHIC.  %to pursue the exploration of the QCD matter phase diagram at centre of mass energies between SPS and LHC.

Last but not least, in our research domain, the progress in theory and physics interpretation is driven by the new insights from experimental data and the precise description of the initial state of the colliding nuclei. In this respect, there will be a very good synergy between the EIC program at BNL, USA and the LHC Heavy Ion program (AA, pA, gamma-A/p) during Run 5. 

In order to answer the previous open questions with the best experimental apparatus, the French heavy ion community fully supports the operation of the LHC with ions (including light and mid-size nuclei) until the end of the machine lifetime~\footnote{The standard 1-month running at the end of the year will also allow the detectors to naturally ``cool off'' after months of very high-luminosity pp operation}, and plans to contribute to the constructions of the new experiments ALICE3 and LHCb upgrade 2.

\section*{ALICE 3}
\label{sec:ALICE3}

In the context of the continuation of the LHC heavy-ion program in Run~5, the end of the designed lifetime of the ALICE detector offers the best and more ambitious opportunities, opening the way for the design of a genuinely dedicated heavy-ion experiment, based on a next-generation detector implementing cutting-edge technologies for vertexing, tracking, particle identification, and readout. Such a new experimental program, unique and complementary to the heavy-ion programs of the other LHC experiments, currently under discussion, is designed to take advantage of the full physics potential of the HL LHC and is being developed under the name of ALICE~3~\cite{ALICE:2022wwr}. The operation of a new, dedicated heavy-ion experiment, complementing the multi-purpose experiments (ATLAS, CMS, and LHCb), would allow to fully exploit the potential of a heavy-ion program at the LHC beyond LS4 and until the end of its lifetime, following the recommendations of the last ESPPU~\cite{EuropeanStrategyforParticlePhysicsPreparatoryGroup:2019qin}.

As a next-generation, dedicated heavy-ion detector at the LHC, ALICE~3 is designed to span a broad and ambitious physics program involving both precision and exploratory measurements, from pp to Pb-Pb collisions, with the final goal of improving our understanding of the mechanisms responsible for the formation and the behaviour of complex hadronic systems and phases, and how the properties of these systems connect to the fundamental parameters of QCD. 

The core of this programme targets measurements aiming at characterising the properties of deconfined hadronic matter and its (self-)interaction with colour-charged particles and radiation: interaction of heavy quarks with the medium, formation and dissociation of bound states inside the medium, emission of electromagnetic radiation from the medium, characterisation of the fundamental QCD symmetries (namely the chiral symmetry) in the medium. 
Other measurements will exploit the specific environment offered by the deconfined phase of hadronic matter to explore the frontiers of QCD in terms of exotic multi-quark and hadronic molecular states, or even search for signals of physics beyond the SM.

\subsection*{Foreseen involvement of the French community in ALICE~3}

The French community aiming at joining the ALICE~3 project currently involves 12 permanent physicists from the IP2I, IPHC and LPSC laboratories.

The French community is among others things interested in the heavy-quark part of the envisaged scientific program of the ALICE~3 project. The main physics goal is to achieve a precise characterisation of the interaction of charm and beauty quarks with the QGP, namely the study of the diffusion, energy-loss and hadronization mechanisms of heavy quarks in the medium and their dependence on the quark mass and their possible sensitivity to the thermal scale (mostly relevant for charm quarks). 
ALICE3 specifics on such front (e.g.~with respect to CMS or LHCb) will be on the explored phase space: the focus will be given to the combination of mid/intermediate rapidities ($|\eta|< 3-4$) with low and intermediate momenta ($0< p_T(HF) < 3$~GeV/c and above).
Key measurements include the angular correlations of charm hadron pairs in a wide rapidity domain and down to low transverse momentum, the anisotropic flow of both charm and beauty mesons and baryons, the yields of multi-charm baryons, and the investigation of the inner structure of heavy-flavour jets. The study of the production of multi-charm baryons is particularly relevant in heavy-ion collisions, confronting directly model predictions: a significant increase of the corresponding yields - due to specific hadronisation mechanisms only available within a deconfined medium - is expected with respect to the vacuum, for which pp collisions can serve as a proxy. Measurements of hadron momentum correlations will also be exploited as a unique tool to investigate specific questions rooted in QCD such as the properties of the hadron-hadron potential between pairs of heavy-flavour hadrons, and set indirect constraints on the nature of exotic hadronic structures.

% Comments taken during the workshop
% \begin{itemize}

% \item Importance de l'existence d'une expérience optimisée HI pour assurer un programme HI au LHC

% \item Complémentarité entre ALICE 3, LHCb et CMS : 3 angles d'attaque à l'étude du QGP

% \item Mettre en avant la volonté de participer aux projets ALICE 3 et upgrade LHCb aussi avec des contributions techniques sur des axes technologiques en adéquation avec les compétences des labos et stratégiques pour les Instituts

% \item Les upgrades LHCb d'intérêt pour les HI s'intègrent dans le programme d'upgrade LHCb Phase II pour la physique des saveurs

% \item Spécificité LHCb : spectroscopie hadronique et caractérisation QGP à l'avant

% \end{itemize}

\subsection*{ALICE~3 French Community Technical Contribution}

The French community in ALICE casts a technical contribution towards the Middle Layers (ML) and more particularly Outer Tracker (OT) sub-systems, building up on many years of expertise on MAPS development (IPHC).
ML and OT are MAPS-based trackers located beyond the beam pipe, hosting the Vertex Detector (e.g. 7~cm $< r[\mathrm{ML}+\mathrm{OT}] <$ 80 cm).
ML and OT cover together mid-rapidity with cylindrical layers ($|\eta| < 1.5$) and intermediate rapidities with disks (1.5~$< |\eta| <$~3--4), assembling about 55~m$^2$ of active silicon. The ML and OT are likely to share a same and unique planar MAPS sensor; such a sensor will be based on the same technological node (65~nm) than the one foreseen for ITS3~\cite{alice2024its3TDR} in Run 4, but the ML/OT sensor will be of a new architecture calling for specific design. The particular stress for such MAPS will be on maintaining a rather small spatial resolution ($< 10~\mu$m) and obtaining a very good time resolution ($<$ 100 ns), coupled with very low material budget ($< 50~\mu$m thickness per layer) and a very low power consumption $\mathcal{O}$(20-40~mW/cm$^2$). in a more aggressive radiation context (e.g. $< 10^{14}~1$-MeV n$_\mathrm{eq}.\mathrm{cm}^{-2}$).\\
The ALICE~3 community in France has expressed ones' intention to work with the technical departments in their institutes on three aspects: i) the CMOS design itself, ii) the middle- and back-end readout electronics and iii) the mechanical integration for mid-rapidity OT staves.
The expected international partners are, among others, Germany, USA, Korea for OT and Italy, CERN for ML.

\section*{LHCb}
\label{sec:LHCb}

A second major upgrade of the LHCb experiment is due to take place to enable full exploitation of the LHCb detector during the HL-LHC for flavour physics, as planned in the European strategy for particle physics. This is the LHCb-U2 project \cite{LHCb:2021glh, LHCb:2018roe}. Using radiation-resistant detection systems, LHCb will take data from pp collisions at instantaneous luminosities of up to $1.5 \times 10^{34}~{\rm cm}^{-2} {\rm s}^{-1}$. This will enable the experiment to reach a total integrated luminosity of 300 fb$^{-1}$, corresponding to unprecedented samples of hadrons of beauty and charm, with a unique sensitivity to physics beyond the SM. LHCb-U2 will also offer several other unique scientific opportunities, including hadron spectroscopy, electroweak precision measurements, the search for the dark sector and the study of heavy ion collisions, while maintaining or exceeding the detector's current performance in pp collisions, and considerably improving its performance in heavy ion collisions.   

In fact, at these luminosities, LHCb should be able to measure the 40 or so, pp collisions that will take place in the same bunch crossing of the LHC. The LHCb-U2 tracker system will be able to reconstruct a large number of tracks per bunch crossing, comparable to the multiplicity of charged particles expected in a collision between heavy ions. For this reason, after the U2 upgrade, LHCb will be able to measure central Pb-Pb collisions and to enrich its scientific program to pursue the exploration of the strongly interacting matter phase diagram at the LHC. The LHCb experiment is also unique in that it covers a pseudorapidity range that is unique to the LHC ($2<\eta<5$), complementing the other LHC experiments: CMS and ATLAS and the future ALICE3 experiment. The scientific program with heavy ions will address two different and complementary physics quests:

\begin{itemize}

\item The origin of the collective-like phenomena observed to date in very light systems such as high-multiplicity pp collisions can be studied in detail with LHCb, by measuring all the intermediate systems between pp and Pb-Pb collisions, such as the p-A, O-O, Ar-Ar or Xe-Xe systems or with other species that can be accelerated in the LHC. Thanks to its excellent temporal resolution, LHCb will be very well placed to study pp collisions at high multiplicity and collisions between light heavy ions at very high luminosity, up to around 10 MHz of the light-ion collision rate. 

\item LHCb is the LHC's emblematic experiment for carrying out a fixed-target physics program at the LHC, at centre-of-mass energies close to those of RHIC, under experimental conditions orthogonal to those of RHIC experiments. Integrated high-luminosity p+A studies on several nuclei will become accessible.  

\end{itemize} 

After LS4, LHCb will have an excellent tracker system for measuring collisions between heavy ions, high-multiplicity p-A and p-p collisions, perfectly suited to measure heavy flavours, quarkonium, prompt dimuons and electroweak bosons. Whereas the production of heavy-quarks is amenable to perturbative QCD and attributed to the initial stages, their kinematic distribution, the correlation with the bulk energy-momentum deposition and their chemical distribution among the different hadrons can be used to infer information about the QGP. 
In particular, their kinetic and chemical thermalisation are expected to be retarded or incomplete. In addition, the heavy quarks are conserved after production. This opens the possibility to study diffusion and hadronisation modifications with respect to equilibrium expectations providing information inaccessible with light-flavour hadrons. 
LHCb has excellent heavy-hadron detection capabilities down to zero transverse momentum and the forward acceptance covers the region where the heavy-quark density and initial energy density varies strongly.  In addition, the study of bound states of two heavy quarks within a meson or a baryon as conventional quarkonium, $B_c$ and $\Xi_{cc}$ and others  are the key channels to characterise the deconfinement transition from QGP to hadrons. These states can be addressed from pp to heavy-ion collisions  with LHCb based on the largest samples in pp collisions as baseline for studies in heavy-ion collisions.

Regarding the study of the thermodynamic properties of the QGP, the LHCb experiment has also unique capabilities. The pseudorapidity coverage allows the measurement of the QCD equation of state properties as a function of the temperature. Thanks to the forward acceptance and the wide coverage, the effective temperature range is about $T_{eff} \in [190, 225]$~MeV~\cite{Mariani:2917083}. Some of the quantities that can be studied as a function of temperature for the first time are the speed of the sound, medium viscosities and quarkonia suppression. 

At high collision energy, the phenomenon of gluon saturation has been predicted to occur in hadrons at small fractional longitudinal momentum. The search for saturation is one of the major motivations for hadron structure measurements at the LHC, in particular at forward rapidity. This is key motivation for the study of Drell--Yan, the study of quarkonium photoproduction and other channels to be explored in ultra-peripheral collisions, or the study of electroweak bosons, where LHCb will provide unique measurements in its rapidity coverage with an excellent momentum resolution. The availability of hard exclusive and inclusive production in hadro- and photoproduction over a broad and highly complementary to the upcoming electron-ion collider kinematic domain using collider and fixed-target mode data complements the hadron structure physics case. 
LHCb allows to detect dimuons down to the charmonium mass region with precision and with very good rejection of secondary vertices thanks to its specialisation for heavy-flavour decays. Multiple processes with dimuon final states are key to understand the mechanisms of isotropisation and chemical equilibration in the initial stages, which remain otherwise unconstrained by experimental measurements~\cite{Coquet:2021gms,Coquet:2023wjk}.
In addition, the photoproduction of charmonium in peripheral and semi-central Pb-Pb collisions will be measured with excellent resolution in transverse momentum and at high statistics. The concomitant measurement of the collision reaction plane with the coherent photoproduction of quarkonia will provide a 2D image of the Pb-Pb collision. 

In addition, the envisaged future project LHCb-Spin~\cite{Aidala:2019pit} for a polarized target will allow unique measurements for spin physics, an important part of hadron structure physics. This program allows the access to complementary channels with respect to the electron-ion collider in the United states.

\subsection*{LHCb Heavy-Ion French Community}
A group of 25 permanent physicists of IJCLab, Irfu, LLR, LPCA and Subatech is already and will continue joining the LHCb collaboration to reinforce the heavy-ion physics working group of the LHCb at the horizon of Run 4 and Run 5. In addition Irfu, LLR and Subatech have expressed their interest to contribute to the construction of a MAPS detector at the Upstream planes, in synergy (same CMOS ASIC) with the Mighty Pixel Tracker detector. This new community will be added to the 29 physicists already working on pp physics in France.

\subsection*{LHCb Heavy-Ion French Community Technical Contribution}

There is no doubt that one of LHCb's most crucial subdetectors for carrying out heavy ion physics with LHCb-U2 is the \emph{Upstream Pixel Tracker}. In order to be able to have a tracker system that would make it possible to reconstruct tracks in high-multiplicity events, the use of MAPS sensors is becoming  a must as a replacement for the silicon strip modules of the U1 Upgrade of the \emph{Upstream Pixel Tracker}. In addition, the high-pseudorapidity acceptance of LHCb's Middle Tracker will also have to be covered by MAPS sensors, as part of the Migthy Pixel Tracker project, to ensure that the tracker system has excellent trace reconstruction performance for high-multiplicity events in the largest pseudorapidities. \emph{Upstream Pixel Tracker} and \emph{Migthy Pixel Tracker} projects share similar requirements and one single MAPS sensor should meet the specifications of both projects. Using the same sensor has practical advantages in terms of simplifying the operation and maintenance of the detector, but that's not all : it considerably reduces the human resources required for its design and assembly. Firstly, in microelectronics, it enables beam testing campaigns and laboratory characterisation of different prototypes. It also simplifies the integration of the sensor into the experiment, because the mechanics of the modules, their assembly with the sensors, the power supply system, the cooling system and the data acquisition and concentration chain naturally become the same for both LHCb sub-projects. 
In addition, the developments around CMOS MAPS are considered as strategical within the French community since they rely on a very strong expertise both at IN2P3 and Irfu and pave the basis for the new detectors (especially for high precision vertexing and tracking) in the high dose environment of the future FCC experiments. Indeed, together with the joint effort between the \emph{Upstream Pixel Tracker} and the \emph{Migthy Pixel Tracker}, the French community can benefit of the synergy already in place between IPHC and Irfu, and integrated within the ECFA-DRD3, around the R\&D on the 65 nm CMOS technology which may become a possible alternative for LHCb with respect to the present developments on more standard technologies. All those joint efforts place the French community in a very favorable position to strengthen both projects and enlarge the present collaboration of the \emph{Upstream Pixel Tracker} project beyond France and China.

We would like to stress that we must not forget that, beyond heavy-ion physics, the LHCb-U2 upgrades (\emph{Upstream Pixel Tracker} and \emph{Mighty Pixel Tracker}, as well as the measurement of the time of flight with resolutions challenging of a few tens of picoseconds by the VELO sub-detectors, the Cherenkov detectors \emph{RICH}, the electromagnetic calorimeter and the time-of-flight detector \emph{TORCH}) are imposed by the need to increase the integrated pp luminosity. From this point of view, the physics of heavy ions becomes a plus and is added in a natural way to all the other physics goals that the LHCb experiment will be able to tackle during Run 5 at the LHC, in particular physics beyond the SM.

\section*{Future Colliders at CERN}
\label{sec:FCC}

The proposed CERN Future Circular Collider (FCC) in its electron-positron $\rm e^+e^-$ (FCC-ee)~\cite{FCC:2018evy} and hadron-hadron (FCC-hh)~\cite{FCC:2018vvp} running modes, will be an outstanding facility for new and/or high precision QCD measurements. 

In particular, the FCC-hh run with Pb-Pb collisions at nucleon-nucleon centre-of-mass energy of $\sqrtsnn = 39$~TeV will produce a deconfined state of QCD matter at energy densities, about 40 GeV/fm$^3$ at nominal times of $\tau_0\approx1$~fm/$c$, never explored before~\cite{FCC-ionsstudygroup:2017glf}. At these energies, about 500 charm quark pairs will be produced in central Pb-Pb collisions, becoming a major ingredient to probe the quark-gluon soup, which will be in a QGP state up to $\tau\approx 15$~fm/$c$. The increased collision energies and integrated luminosities, about one order of magnitude higher than the LHC each, will significantly increase the number of hard and heavy particles produced and usable to ``tomographically'' probe the plasma. For the first time, the top-quark (and even the Higgs boson) will be produced in large quantities to uniquely probe and characterise the QGP, in addition to many other observables like multi-TeV jets, photons, electroweak bosons, and charm and bottom quarks. The extremely high density and high energy of the photon field in ultraperipheral heavy-ion collisions will provide novel searches for new physics (e.g.\ axion-like or graviton-like particles) using photon-photon collisions~\cite{Shao:2022cly}. Last but not least, extremely low gluon fractional momenta (down to $x\approx 10^{-7}$) in the nuclear parton densities will be explored with perturbative probes in proton-lead (p-Pb) collisions at $\sqrtsnn = 63$~TeV.

The FCC-ee will produce extraordinarily large samples of Z ($8\cdot10^{12}$), W ($5\cdot10^8$) and Higgs ($2\cdot 10^6$) bosons that, in their dominant hadronic decays, will provide unique sets of self-flavour-tagged (via tag-and-probe techniques) light- and heavy-quark and gluon jets allowing their precise study and characterisation under extremely clean conditions. 
High-precision QCD measurements~\cite{Skands:2016bxb}, which are crucial for indirect searches of physics beyond the SM at FCC-ee, will be at reach such as determinations of the strong coupling constant with permille uncertainties, ultrapure samples of flavour-tagged jets, accurate energy-angle analyses of hadrons inside jets to constrain high-order (fixed and logarithmic) perturbative corrections of parton showers via Lund Jet plane techniques, multidifferential studies of final hadrons for very precise determinations of fragmentation functions, pristine experimental conditions to investigate non-perturbative phenomena (colour reconnection, hadronization, final-state interactions,...), as well as the production of particularly exotic QCD bound states (from ultrarare hadronic decays of $8\cdot10^{12}$ Z bosons).

\begin{comment}
The French QCD community strongly supports the plans of building a CERN circular $\rm e^+e^-$  collider at centre of mass energy between the Z boson mass and twice the mass of the top quark to carry out multiple precise studies of the strong interaction. Such a collider should start the commissioning after the end of the LHC program. After the realisation of the FCC-ee physics program ---and as soon as the R\&D activities, mainly of high-field magnetic dipoles,  will be finalised--- the FCC tunnel will host an energy-frontier pp and heavy-ion collider with unique capabilities for QCD and QGP studies. 
The community acknowledges the large temporal gap between HL-LHC and FCC-hh without hadron-hadron or DIS colliders, apart from the electron-ion collider program in the USA for hadron structure.  Hence, we encourage the investment in accelerator technology and its close monitoring  for the use for hadron structure and QGP physics at high collision energy as secondary projects in parallel to the FCC-ee program. In this view of a large temporal gap, it is a natural consequence that the community would be interested in having FCC-hh after LHC. The needed parton distribution measurements for the full exploitation of the FCC-hh could be provided by a prolongation of the LHC program with the LHeC.
\end{comment} 

The French QCD community strongly supports the plans of building a CERN circular collider. The community acknowledges the scientific interest of multiples high precision measurements of the strong interaction with FCC-ee at center of mass energy between the Z boson mass and twice the mass of the top quark. 
Nevertheless, a FCC-hh will give a larger and complete program to study the QGP at unprecedented energy densities. If a FCC-ee collider comes just after the HL-LHC operation, then we will have a large temporal gap between HL-LHC and FCC-hh without hadron-hadron or DIS colliders, apart from the electron-ion collider program in the USA for hadron structure.
In the scenario of having a FCC-hh first, the gap needed to finalizing the necessary R\&D on high-field magnets could be covered by a LHeC machine. The LHeC will provide  the parton distribution measurements needed for the full exploitation of the FCC-hh data. In addition, the opportunities to explore the high-baryon density region of the QCD phase diagram  at fixed-target energies with dedicated experiments
 complementary to FAIR with the CERN accelerator complex should be explored.

\section*{Executive summary}
\label{sec:summary}

In preparation of the LHC Run 5 program and future collider projects at CERN, the French QCD community : 
\begin{itemize}
\item Advocates for the commitment, of present LHC French Community, to the full exploitation of the physics program with ions for the understanding of QCD processes and the QCD phase diagram (and, more broadly, of the collective behaviour of non-Abelian quantum field theories), scanning a broad range of species and systems, and during the full lifetime of the HL-LHC operation.

\item Advocates for the engagement of the French community  in the  two experimental projects, to fully exploit the ion physics program during the LHC Runs 5: ALICE3 and LHCb. Keeping also in mind the valuable complementarity for heavy-ion measurements provided by the general-purpose detectors ATLAS and CMS.  

\item Highlights the strong complementarity of the LHC Run 5 and the ion program at the EIC.

\item Supports the next generation machine FCC-ee and FCC-hh with very broad, different and complementary QCD physics programs.

\item Supports the theoretical developments required for the full interpretation of the collected data. 
\end{itemize}

%\break
\printbibliography

\end{document}